\documentstyle[prl,twocolumn,aps,epsf]{revtex}

\begin{document}

\title{Spin bath-mediated decoherence in superconductors}

\author{N. V. Prokof'ev$^{1,2}$  and P. C. E. Stamp$^{2,3}$}
\address{$^{1}$ Physics Dept., Hasbrouck Lab., University of
Massachussetts, Amherst, MA 01003, USA  \\
$^{2}$ Institute for Theoretical Physics, University of California at
Santa Barbara, CA 93106 \\
$^{3}$ Spinoza Institute, Minnaert Bld., University of Utrecht, 
Leuvenlaan 4, 3584 CE Utrecht, Netherlands}

\maketitle

\begin{abstract}

We consider a SQUID tunneling between 2 nearly degenerate flux states. 
Decoherence caused by paramagnetic and nuclear spins 
in the low-$T$ limit is shown to be much stronger than that from 
electronic excitations. 
The decoherence time $\tau_{\phi}$
is determined by the linewidth $E_o$ of spin bath states, which can be reduced
by a correct choice of ring geometry and isotopic purification. 
$E_o$ can be measured in
either field sweep or microwave absorption experiments, allowing both a test
of the theory and design control.

\end{abstract}

\bigskip

\pacs{PACS numbers: }

The fundamental importance of decoherence in Nature has been
emphasized in fields ranging from quantum gravity and
measurement theory to quantum computing and mesoscopic physics.
Serious analysis of the {\it mechanisms} controlling
decoherence began with work on dissipative tunneling and on 
``Macroscopic Quantum Coherence''  
(MQC) in SQUIDs \cite{ajl87,ajl80}. 
To see MQC in SQUIDs requires 
$\tau_{\phi} (T) \vert \Delta \vert \gg 2\pi \hbar $, where $\tau_{\phi}$
is a ``decoherence time'', and $\Delta $ the 
tunneling matrix element. However, despite the successful
observation of macroscopic tunneling (MQT) in SQUIDs \cite{voss},  
MQC has not been found. 
The unexpected low-$T$ saturation of $\tau_{\phi}(T)$ in 
conductors
\cite{mohanty,golubov,zawad}, indicates 
we may not understand decoherence even in metallic systems
(let alone in ``qubits'' or quantum computers!).

Our thesis herein is that at
low $T$ in SQUIDs, $\tau_{\phi}(T)$ is controlled not by
electronic or ``oscillator bath'' environments
\cite{ajl80,amb85,larkin}, but by a ``spin bath'' \cite{rpp}
of localized modes, including nuclear and paramagnetic spins
as well as charge defects. Although spin bath environments have
received considerable attention in nanomagnets \cite{ps,ohm98}
(and now in mesoscopic conductors \cite{zawad}) their discussion
for SQUID tunneling has been only sporadic \cite{rpp,sta88}.
Here we show how they strongly suppress MQC (but usually with rather 
weak effects on MQT).
Our results should allow experimentalists to both
parametrize the spin bath effects, and quantitatively test the theory.

\vspace{3mm}

1. {\it Effective  Hamiltonian}. 
Ignoring electronic and
electromagnetic dissipation mechanisms
\cite{ajl80,amb85,larkin}, the "bare" DC SQUID Hamiltonian for flux 
tunneling is
\begin{equation}
H_0={p_{\phi}^2 \over 2 C} +U_0\left[ {(\phi-\phi_x)^2 \over 2} +
g \cos \phi \right] \;,
\label{1}
\end{equation}
where $C$ is the junction capacitance, 
$\phi = 2\pi \Phi /\Phi_0$ with $\Phi_0$ the flux quantum, 
$p_{\phi} = -i\hbar \partial
/\partial \Phi$,  
and $\phi_x= 2\pi \Phi_x /\Phi_0$, where 
$\Phi_x$ is the externally applied flux. Then 
$U_0=\Phi_0^2/4\pi^2 L$ 
and $g=2\pi LI_c/\Phi_0$, for a ring inductance $L$ and junction critical  
current $I_c$. 
In MQC or qubit designs
$\phi_x \ll 1$ and $\phi$ 
tunnels between the two lowest wells, centred at $\phi =
\phi_{\pm} = \pm \phi_m$. One also assumes  
$k_BT < \hbar \Omega_0 /2\pi$, where
$\Omega_0$ is the Josephson plasma frequency of small
oscillations in these wells. The system then truncates 
\cite{ajl87,ajl80,amb85} to a
two-level Hamiltonian $H_0 = \xi_o(t) \tau_z + \Delta \tau_x$, where
$\vec{\tau }$ is a Pauli vector, $\xi_o(t) \sim U_0 \phi_m
\phi_x(t) $ is a bias which can be varied in time, 
and current designs have $\vert \Delta /\hbar \vert$ as high
as $\sim 1~GHz$.

Consider now a set of $N$ two-level systems 
$\{ \vec{\sigma}_k \} \equiv \{ \vec{s}_k,
\vec{I}_k \}$, representing  paramagnetic ($ \{ \vec{s}_k \}$) and 
nuclear ($\{ \vec{I}_k \}$) spins \cite{spinhalf}, at positions
$\{ {\bf r}_k \}$.
Those in the SQUID couple  
to the conduction electron spin density $ \vec{s}_{\rm cond} ({\bf r}) $ 
via 
\begin{equation} 
H_{sp}  = \sum_k J_k \vec{\sigma}_k \cdot
\vec{s}_{\rm cond} ({\bf r}_k) 
+ \gamma_k \vec{\sigma}_k \cdot \vec{B}({\bf r}_k)
\;, \label{3}  
\end{equation}
Spins in the substrate couple to the flux via the 2nd term.
The $\{ \gamma_k \} \equiv g_k \mu_k$ are spin
magnetic moments, and the $\{ J_k \}$ represent 
electronic exchange for 
the $\{ \vec{s}_k \}$, and hyperfine
coupling for the $\{ \vec{I}_k \}$. 
All $\{ \vec{s}_k \}$ are assumed paramagnetic \cite{14},
with Kondo energies $k_BT_K \sim (JE_F)^{1/2}
e^{-1/JN(0)} \ll k_BT$,
where $N(0)$ is the Fermi-level density of states. 
The field $\vec{B}({\bf r}) = \vec{B}_{x} + \vec{B}_{\phi}({\bf
r})+\sum_k \vec{b}_{k}({\bf r})$, where $\vec{B}_{x}$ is the
external field, $\vec{B}_{\phi}({\bf r})$ comes from the SQUID 
supercurrent, and 
\begin{eqnarray}
\vec{b}_{k}({\bf r}) &=& \gamma_k \bigg[ {8\pi \over 3} 
\vec{\sigma}_k \delta ({\bf r}-{\bf r}_k) \nonumber \\
& + &  {\vec{\sigma}_k \over \vert {\bf r}-{\bf r}_k \vert^3 } - 
{ 3 ({\bf r}-{\bf r}_k) \cdot \vec{\sigma}_k  \over 
\vert {\bf r} - {\bf r}_k \vert^5 }
({\bf r}-{\bf r}_k)  \bigg] 
\;, \label{4}  
\end{eqnarray}
is the dipolar field from $\vec{\sigma}_k$ 
if $\vert {\bf r} - {\bf r}_k \vert \ll \lambda_s$, the superconducting 
penetration depth \cite{Suhl}.

>From $H = H_o + H_{sp}$ we now derive a low-energy Hamiltonian, using the usual 
instanton method \cite{ajl87} wherein
tunneling of $\phi$ occurs
in imaginary time  $\tau$. A transition at $\tau =0$ gives a
variation $\phi (\tau ) = \phi_m f(\tau )$, with $f(\tau ) \sim
\pm \tan^{-1}( e^{\Omega_0 \tau } )$.
Consider now the local fields $\vec{\omega}_k = \gamma_k
\vec{B} ( {\bf r}_k)$ acting on the $\{ \vec{\sigma}_k \}$,
which during the instanton evolve like:
\begin{equation}
\vec{\omega}_k (\tau) = \omega_k^{\perp} \vec{m}_k  +
\omega_k^{\parallel} \vec{l}_k (\tau )  
\;, \label{5}  
\end{equation}
(cf. Fig.~1), where $\vert \vec{m}_k \vert =1 $ and $ \vec{l}_k (\tau )$
evolves from $\vec{l}_k^{\pm}$ to $-\vec{l}_k^{\pm}$ during the
instanton; $\vert \vec{l}_k^{\pm} \vert =1 $ at the end-points. 
Then, defining
$\vec{B}_{\phi }^{\pm} ({\bf r}_k) =
[\vec{B}_{\phi = +\phi_m}({\bf r}_k) 
\pm \vec{B}_{\phi = -\phi_m}({\bf r}_k)]$, 
\begin{eqnarray}
& & \omega_k^{\perp} \vec{m}_k = \gamma_k \{ 
\vec{B}_x + \sum_j \vec{b}_j ({\bf r}_k ) +{1 \over 2} 
\vec{B}_{\phi }^{+}({\bf r}_k) \} 
\label{6} \\ 
& &  \omega_k^{\parallel} = \gamma_k {1 \over 2}
\vert  \vec{B}_{\phi }^{-}({\bf r}_k)  \vert \; \equiv \; 
{1 \over 2} \gamma_k \delta B_k
\label{7}  
\end{eqnarray}
The time varying $\omega_k^{\parallel} \vec{l}_k
(\tau )$ causes transitions of $\vec{\sigma}_k$; writing $\vert \sigma_k^+> =
T_k^{\pm} \vert \sigma_k^- >$, which defines a transition matrix
$T_k^{\pm}$ for the passage $\vec{l}_k^- \to \vec{l}_k^+$, we
have in general
\begin{equation}
T_k^{\pm} = \exp \left\{ -{1 \over \hbar} \int_-^+ d\tau 
\omega_k^{\parallel} \vec{l}_k (\tau ) \cdot \vec{\sigma}_k
\right\} \equiv e^{-i\varphi_k +\vec{\alpha}_k \cdot \vec{\sigma}_k }
\;, \label{8}
\end{equation} 
operating on $\vec{\sigma}_k$, where $\varphi_k,\alpha_k$
are typically complex \cite{rpp}. In the long
interval (over times $\sim \hbar/ \Delta $) between
instantons, $\vec{\sigma}_k$  sits in a static field
$\vec{\omega}_k^{\pm} \equiv \vec{\omega}_k(\vec{l}_k = \vec{l}_k^{\pm})$; 
but during instantons, $ T_k^{\pm}$
operates. This gives immediately the diagonal and non-diagonal terms in
an effective Hamiltonian of 
``Central Spin'' form \cite{rpp,ps}, valid at energy scales 
$\ll \hbar \Omega_o$: 
\begin{eqnarray}
H_{\rm eff} = & & \left[ \Delta \hat{\tau}_+ e^{-i\sum_k
\vec{\alpha_k} \cdot \vec{\sigma}_k } + H.c. \right] \nonumber \\
&& + \sum_k \left( \omega_k^{\perp} \vec{\sigma}_k \cdot \vec{m}_k   +
\hat{\tau}_z \omega_k^{\parallel} \vec{\sigma}_k \cdot \vec{l}_k \right)+
\xi_o (t) \hat{\tau}_z  
\;, \label{9}  
\end{eqnarray}
where the $ \varphi_k$ are absorbed into the physical 
$\Delta $. Eigenstates $\vert 
\sigma \rangle$ of 
$\hat{\tau}_z$, with $\sigma = \pm 1$ for $\tau_z = \uparrow, \downarrow$, 
are converted 
to a pair of $2^N$-fold multiplets of coupled SQUID/spin bath states, with 
linewidth  $E_0^2=\sum_k (\omega_k^{\parallel})^2 $
and normalised 
distributions
$W_{\sigma}(\xi) = (2/\pi E_0^2)^{1/2} 
e^{-2 (\xi -\sigma \xi_o)^2/E_o^2}$, in energy $\xi$. The parameter $E_o$ will
be crucial to decoherence- note it only depends on the 
{\it change} $\delta B_k$ 
in field in (\ref{7}), and is much easier to determine than the total field
$\vec{B}({\bf r}_k)$.  

The above derivation ignores the slow spin diffusion
and spin-lattice relaxation in the spin bath (which cause the vectors
$\vec{l}_k$,$\vec{m}_k$ to become dynamic variables). This 
occurs on timescales $\sim \mu s$ or longer, ie., $\gg \hbar / \Delta$, and is not 
relevant to decoherence in the present problem.

\vspace{3mm}

2. {\it Example}: 
The parameters $\omega_k^{\perp}$, 
$\omega_k^{\parallel}$ are determined from 
the fields $\vec{B}_{\phi}^{\pm} 
({\bf r}_k)$ (ie., knowing the 
supercurrent distributions corresponding to $\phi = \pm \phi_m$), once 
we know $\{ \gamma_k \}$ and $\{ {\bf r}_k \}$ for all relevant
nuclear and paramagnetic spins. We therefore assume homogeneous concentrations 
$x_r$, $x_J$, and $x_s$ of paramagnetic spins in the ring, junction, and 
substrate respectively, for the geometry in Fig.~2, 
as well as a single nuclear species, with one 
nucleus per unit cell. Simple magnetostatics gives the results in
Table I for $\omega_k^{\perp}$ and $\omega_k^{\parallel}$.
The vector $\vec{\alpha}_k$ depends
on the detailed path followed by $\vec{l}_k(\tau)$ during tunneling. Since 
$\vec{l}_k(\tau)$ changes very quickly (on a timescale $\Omega_o^{-1}$), 
from time-dependent perturbation theory $\vert \vec{\alpha}_k \vert
\sim \omega_k^{\parallel}/\hbar \Omega_o$, and the mean number of bath spins
flipping per transition is $\lambda \sim {1 \over 2} \sum_k 
\vert \vec{\alpha}_k \vert^2 \sim  E_o^2/\hbar^2 \Omega_0^2$ 
(with an associated decoherence rate 
$\lambda \Delta/ \hbar  \sim \Delta E_o^2/\hbar^3 \Omega_o^2 $). Thus 
$\lambda \ll 1$ unless $E_o$ is unreasonably large.

The dependence on sample size is the most striking result in Table I 
and Fig. 2, 
the contributions of all spins 
to $E_o$ increasing rapidly as $R$ {\it decreases}-
the increase in the number of spins with $R$
is more than offset by the weaker fields. 
The contribution from the substrate can 
easily be reduced by making it very thin and surrounding 
it with $^4He$,
and one may also reduce $E_o$ by reducing $d$,$h$,$x$, 
and $\phi_m$ \cite{phi_m}.

\vspace{3mm}

3. {\it SQUID  dynamics}: A proper calculation of the SQUID dynamics requires 
adding to (\ref{9}) a coupling to an oscillator bath, representing 
electrons, photons, and phonons. In the absence of a spin bath, the 
dominant electronic contribution gives a decoherence rate \cite{amb85,larkin} 
$\Gamma_{\phi }^{e} = 16\pi \phi_m^2 k_BT /Re^2 = \tilde{\alpha}_e kT/\hbar$ where 
$\tilde{\alpha}_s = (16\pi \phi_m^2 \hbar \Omega_o/ E_c) Q^{-1}$,
$Q(T)$ is the SQUID $Q$-factor, and $E_c$ the junction charging energy. 
In a simple RCSJ model 
with shunt resistance $R_s$ and junction resistance $R_j$ 
one has $R^{-1} = R_s^{-1} + R_j^{-1}$  (and 
$R_j(T) \sim R_o (1+e^{\Delta_{BCS}/T})/2$, where $\Delta_{BCS}$ is the 
superconducting gap and $R_o$ the normal junction resistance), so that 
$Q$ and $\tilde{\alpha}_s$ are $T$-independent at low $T$. 

Let us now calculate the SQUID dynamics for the Hamiltonian (\ref{9}),
and thence the spin bath contribution to $\tau_{\phi}^{-1}$; 
we then compare this to the electronic contribution.
Assuming $\lambda \ll 1$ (see above) we drop the  
$\{\alpha_k \}$ from (\ref{9}). 
Since $\omega_k^{\parallel}, \omega_k^{\perp} \ll
\Delta$, 
we can treat these couplings  
perturbatively; and a quick check of the numbers in Table I shows 
$\omega_k^{\parallel} \ll \omega_k^{\perp}$ almost always.
Since in general $\vec{m}$ and $\vec{l}$ are neither parallel
nor perpendicular, we choose $\vec{m}$ as the 
spin quantization axis $\hat{\vec{z}}$.
The component of $\vec{l}$ 
parallel to $\vec{m}$ is dealt with by redefining 
$\xi \to \xi + \sum_k \omega_k^{\parallel} 
\cos (\hat{\vec{l} \vec{m}} ) \sigma_k^z$, which now depends on the 
spin bath state. Because  
$\vec{l}_x$ has a transverse component, 
whenever the SQUID state changes the
coupling term $\sum_k  \omega_k^{\parallel} \vec{\sigma}_k^x \: \tau_z$ 
forces environmental spins to precess in a new local magnetic field, which 
can be viewed quantum mechanically as SQUID-induced transitions 
between the environmental 
states \cite{rpp}. To quantify this effect we 
calculate the
time correlation function $P_{\uparrow \uparrow} (t)$, the probability
\cite{ajl87}  that $\tau_z = 1$ at time $t$ (ie., $\phi = +\phi_m$)
if it was $1$  at
$t=0$, after integrating out the spin bath.
An instanton expansion gives (assuming 
 $\xi \ll \vert \Delta \vert$):
\begin{equation}
P_{\uparrow \uparrow} = \sum_{nm} \vert i\Delta \vert^{2(n+m)}
\prod_{a=1}^{2n} dt_a \prod_{b=1}^{2m}dt_b'  
F \left[ \tau_z(t), \tau_z(t') \right]
\label{10}
\end{equation}
summed over ``outgoing'' and ``return'' paths $\tau_z(t),\tau_z'(t)$.
The influence functional $F$ \cite{feyv63} given from (\ref{9}), assuming
$\vert \vec{\alpha}_k \vert = 0$, $\omega_k^{\perp} \ll k_BT$, and 
$\omega_k^{\parallel} \ll \omega_k^{\perp}$, is
\begin{equation}
\ln F= -\sum_k { (\omega_k^{\parallel} )^2 \over 8\hbar^2 }
\vert \int_0^t  ds  e^{{i \over \hbar} \omega_k^{\perp} s } 
\big[ \tau_z(s)- \tau_z'(s) \big] \vert ^2 
 \label{11}
\end{equation}
(assuming an initial thermal environmental state).
We distinguish 
2 regimes:

(a) Strong decoherence regime ($E_0 \gg \Delta$): Here $F$ is
negligible if instanton/anti-instanton pairs are separated by
times $>\hbar/E_o$, and we find immediately that 
$F \sim \exp [-E_o^2(t_i-t_{i+1})^2/2\hbar^2 ]$,
implying $P_{\uparrow \uparrow}(t) \sim (1/2)[1+e^{-t/\tau_{R}}]$, with the  
rate 
\begin{equation}
\tau^{-1}_R = {2\Delta^2  \over \hbar^2} 
\int_0^{\infty} dt e^{-E_o^2t^2/2\hbar^2} =
\sqrt{ 2\pi } {\Delta^2 \over \hbar E_o  } \;,
\label{12}
\end{equation}
ie., completely incoherent quantum relaxation.

(b) Weak decoherence regime ($E_o \ll \Delta$): This problem is 
easily solved by 
going to the basis of eigenstates of $\hat{\tau}_x$.
Their associated spin multiplets are widely separated in energy, so real 
transitions between them are impossible, and a perturbation expansion in 
$\omega_k^{\parallel}/\Delta$ gives 
\begin{equation}
H_{\rm eff} \sim \left[ \Delta + \sum_{kk'} {\omega_k^{\parallel} 
\omega_{k'}^{\parallel} \over 2 \Delta} \hat{\sigma}_k^x \hat{\sigma}_{k'}^x
\right] \hat{\tau}_x + \sum_k \omega_k^{\perp} \hat{\sigma}_k^z  
\label{12a}
\end{equation}
plus terms $\sim O((\omega_k^{\parallel})^4/\Delta^3)$. Assuming $E_o \ll k_BT$
(so all states of a multiplet have equal thermal weight), 
we find that 
\begin{equation} 
\tau_{\phi}^{-1} \sim \sum_{kk'}\omega_k^{\parallel} \omega_{k'}^{\parallel}
/8 \Delta \;\approx \;\; E_o^2 / 8\hbar \Delta
\label{tau-phi}
\end{equation}
ie., roughly $ \Delta \tau_{\phi}/2\pi \hbar   \sim (\Delta/E_o)^2$ 
oscillations survive
before phase decoherence sets in.
We emphasize that {\it no energy relaxation from 
$\vert 1 \rangle$ to $\vert 0 \rangle$ is involved here}- this 
requires coupling to 
electronic excitations \cite{ajl87,ajl80,amb85,larkin}. 

Comparing now electronic and spin bath decoherence rates, we see  
a crossover from electronic-dominated to spin bath-dominated
decoherence around a temperature 
$T_c \sim (2 \pi)^{1/2} \Delta^2/k_B \tilde{\alpha}_s E_o$
(strong decoherence regime), or $T_c \sim E_o^2/8k_B \tilde{\alpha}_s \Delta$
(weak decoherence regime). In the former case coherence will never be seen at 
any $T$; in the latter case the decoherence time will saturate at the 
value in (\ref{tau-phi}), below $T_c$. 
In both cases the low $T$ decoherence is controlled by 
the spin bath. We emphasize, on the other hand, that the spin bath  
will be almost invisible in MQT experiments- it causes litle dissipation, and
adds to the tunneling exponent a factor $\delta S \sim \pi E_o/\hbar \Omega_o$,
so typically $\delta S \ll 1$ (and is $T$-independent!).

\vspace{3mm}

4. {\it Connection to Experiments}:
The concentration, type, and location of the paramagnetic
impurities in the sample will usually be very uncertain. 
Luckily, (\ref{12}) 
and (\ref{12a}) show we only need to know $E_0$ to parametrise the spin bath 
effects. In both 
regimes it can be determined by dynamic "fast passage"
resonant tunneling experiments, from which one can extract 
$W(\epsilon) = W_+(\epsilon) + W_-(\epsilon)$ by 
inverting the data \cite{lukens}.
In the weak decoherence regime we can also look directly 
at the lineshape via microwave absorption between $\vert 0 \rangle$ 
and $\vert 1 \rangle$ manifolds. 

What this means is
that (i) we may characterise the spin bath, extracting $E_o$,
via well-established experimental
techniques, and then (ii) test the theory herein by comparing
the predictions for $P_{\uparrow \uparrow}(t)$ in 
(\ref{12}) and (\ref{tau-phi}) with 
experiment. Note, incidentally, that just as in the nanomagnetic case
\cite{rpp,ps,ohm98}, wide $T$-independent resonant peaks in the sweep experiments are circumstantial evidence for a spin bath-mediated 
mechanism (oscillator bath-mediated relaxation rates are $T$-dependent and 
typically {\it increase}
as one moves further from resonance). Such peaks (of width $\sim 0.4~K$) were
seen in recent experiments \cite{lukens}, indicating a value 
$E_o \sim 0.2~K$ for this particular SQUID. If one is to see MQC, or to make 
superconducting qubits, $E_o$ must be reduced by at least $10^2$. Ways 
to do this were indicated by our example- one wants very pure 
rings (including even isotopic purification if possible) with large
$R$, small $h$, small junctions, thin substrates, and small 
$\phi_m$ \cite{phi_m}.

In this paper we have shown how the low-$T$ decoherence time in a SQUID
must saturate at a value controlled by coupling to the spin bath. 
We thank the Institute for Theoretical Physics in Santa Barbara, where some 
of this work was done, and grant INTAS-2124 from the European community.

\bigskip

{\bf Figures}

\vspace{3mm}

{\bf Fig.~1} (a) Evolution in imaginary time $\tau$ of the field
$\vec{\omega}_k$ acting on $\vec{\sigma}_k$, during tunneling of
$\phi$ from $-\phi_m$ to $+\phi_m$. The field begins at $\vec{\omega}_k^-$
and end at $\vec{\omega}_k^+$. (b) A typical ``path'' (in the
path integral sense) for $\tau_z(t)$, as the SQUID tunnels
between $\vert -\phi_m >$ and $\vert \phi_m >$. The transition
occurs in a ``bounce time'' $\sim \Omega_0^{-1}$; the time
between transitions $\sim 2 \pi \hbar / \Delta$.

%

\vspace{3mm}

{\bf Fig.~2} A model DC SQUID ring, of height $h$, radius $R$, on a
substrate of volume $R^3$. 
The junction is a weak link of length $d$ and radius $r$. We also
show, for this example, the various contributions to $E_o$ as a function of 
$R$, assuming $h = 1~\mu m$, $r = 50~nm$, $d = 200~nm$, and an 
insulating substrate
with a concentration $x_S = 10^{-5}$ of 
paramagnetic impurities. We suppose ring and junction are 
made from $Al$ (with $\gamma_N^{Al} = 11.094 MHz/T$, and $I = 5/2$), with 
concentrations $x_J, x_r = 10^{-6}$  paramagnetic impurities. For definiteness
we assume all paramgnetic impurities have spin $s=1$, $\gamma_s = 2 \mu_B =
14 GHz/T$.


\begin{onecolumn}
\begin{table}
\setdec 0.0
\begin{tabular}{cccc}
~~~~ & Ring & Junction & Substrate \\ 
\tableline
$N_N$ &  $4Rh\lambda_L n_N$ & $\pi r^2 d n_N$ & $\sim R^3n_N$
\\ 
$N_s$ & $x_J N_N$ & $x_JN_N$ & $x_sN_N$  
\\
$\delta B_k$ & $\Phi_0/R^2$ & $\Phi_0/(Rr)$ & $\Phi_0/R^2$
\\
$\omega_k^{\parallel}$ & $\gamma_k  \delta B_k/2$ & $\gamma_k \delta
B_k/2$ & $\gamma_k \delta B_k/2$
\\
$E_0^N$ & $I\gamma_N \Phi_0
(\lambda_Lhn_N)^{1/2}/R^{3/2}$ &$I\gamma_N \Phi_0
(\pi d n_N)^{1/2}/2R$ & $I\gamma_N \Phi_0 n_N^{1/2}/2R^{1/2}$   
\\
$E_0^s$ & $s\gamma_s \Phi_0
(x_J \lambda_L h n_N)^{1/2}/R^{3/2}$ & $s\gamma_s \Phi_0
(x_J \pi d n_N)^{1/2}/2R$ & $s\gamma_s \Phi_0 (x_s n_N)^{1/2}/2R^{1/2}$   
\\
$\omega_N^{\perp}$ & ~~~ & $\gamma_N [\Phi_0/R^2 +
(\mu_o/4 \pi)(\gamma_N+\gamma_s x_{J,s}) C_N n_N ]$ & ~~~
\\
$\omega_s^{\perp}$ & ~~~ & $\gamma_s [\Phi_0/R^2 +
(\mu_o/4 \pi)(\gamma_N+\gamma_s x_{J,s})C_N n_N ]$ & ~~~ 
\\
$\vert \vec{\alpha}_k \vert$ & $\gamma_k \Phi_o/2 R^2 \hbar \Omega_o$ & 
$\gamma_k \Phi_o/2 Rr \hbar \Omega_o$ & $\gamma_k \Phi_o/2R^2 \hbar \Omega_o$ \\
\end{tabular}
\caption{
Parameters for the SQUID in Fig.~2, for nuclear (N) and
paramagnetic (S) spins in the bulk ring, the junction, and the
substrate. $N_N$ and $N_s$ count all spins within
a penetration depth $\lambda_L$ of the surface - in the
ring we assume $h > \lambda_L >r $ (if $\lambda_L >h$,
then substitute $h$ instead of $\lambda$). We assume
$B_x \sim \Phi_0/R^2$. 
The number density $n_N$ of nuclear spins $\{ \vec{I}_k \}$ (with 
$\vert \vec{I}_k \vert = I$) is 
$\sim 1/a_0^3$, where $a_0$ is the lattice parameter; $x_J$ and
$x_s$ are paramagnetic impurity concentrations per site (with 
$\vert \vec{s}_k \vert = s$) in SQUID
and substrate. Values for $\omega^{\perp}$ for ring and substrate spins
(left blank in the Table) are the same as for the junction; $C_N \sim 5-10$ 
is a geometrical factor describing the effective number of 
nearest neighbour spins. 
Finally, $E_0^2=\sum_k (\omega_k^{\parallel})^2$,
so $E_0^s \sim N_s^{1/2} \omega_s^{\parallel}$ and  $E_0^N \sim
N_N^{1/2} \omega_N^{\parallel}$ (see text).  }
\end{table}
\end{onecolumn}


\begin{thebibliography}{99}
\bibitem{spinhalf} For the simple Zeeman coupling in (\ref{3}) 
the only effect of having spin $> 1/2$ is to multiply $E_0$ by $2I$ (nuclear 
spins) or $2s$ (paramagnetic spins).
\bibitem{14} If $k_BT_K \sim \Delta_{BCS}$, the BCS energy gap, 
then we must also include 
mid-gap states in the calculations - see Ref.~\cite{sta88}.
We ignore this possibility here.
\end{thebibliography}
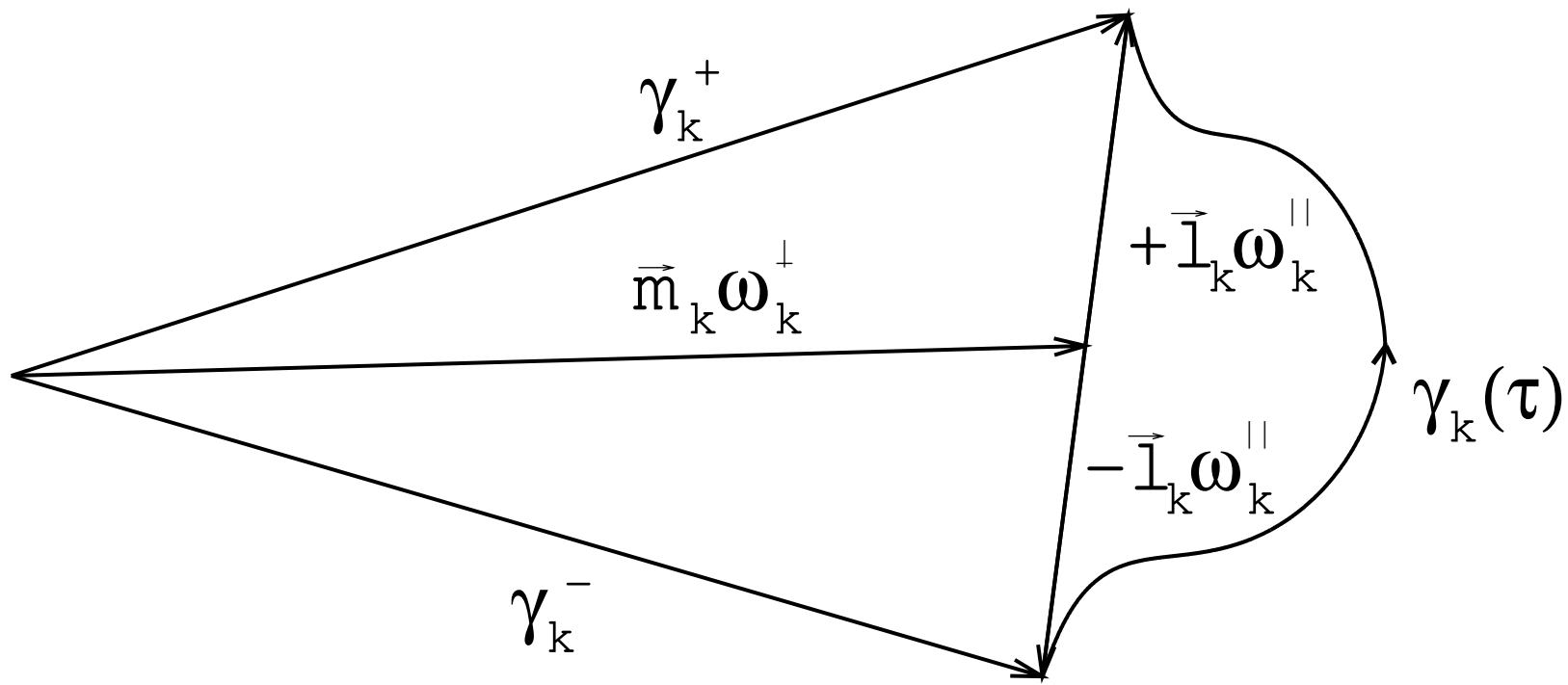
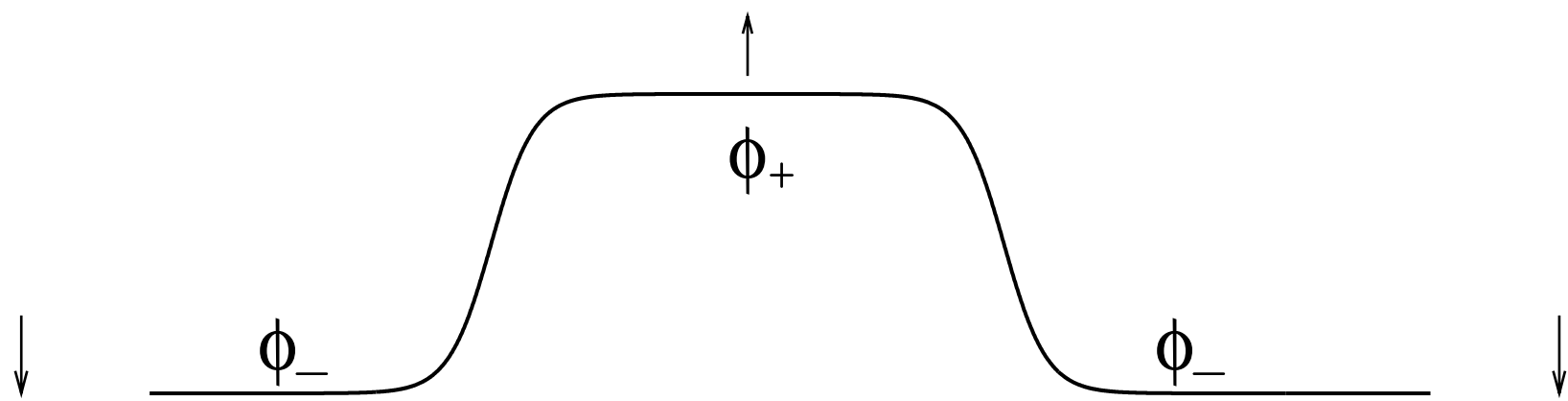
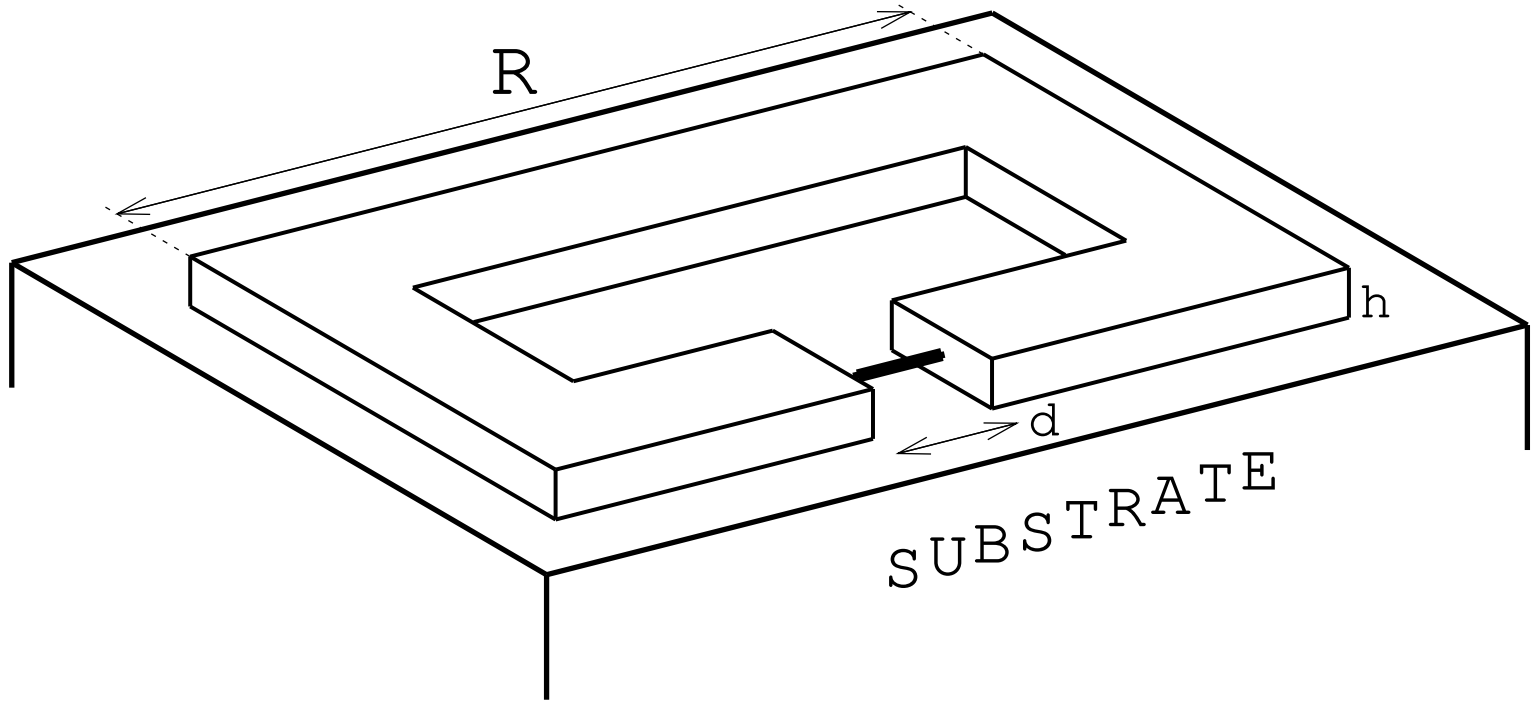
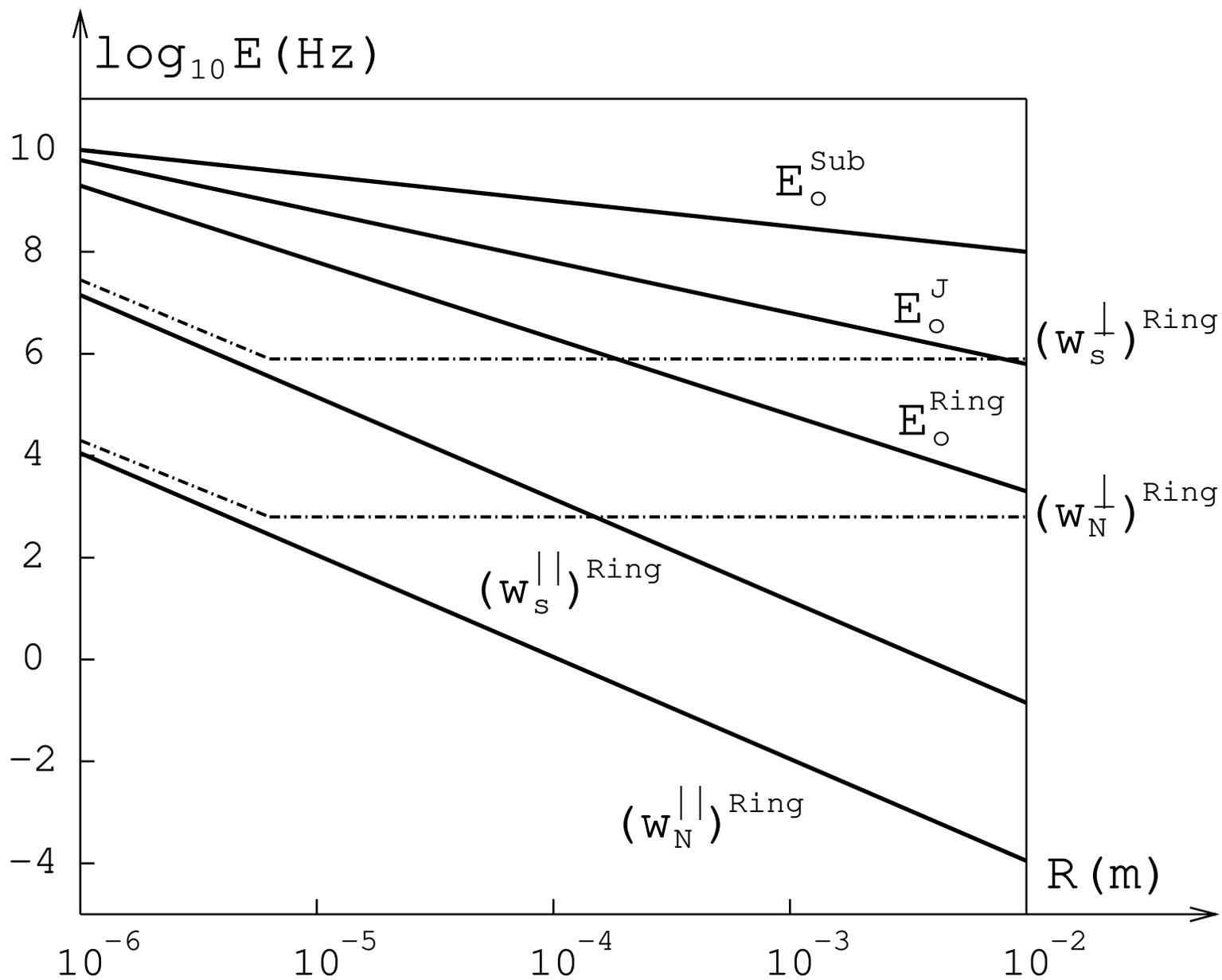
\end{document}